\magnification=1200 \vsize=25truecm \hsize=16truecm \baselineskip=0.6truecm
\parindent=1truecm \nopagenumbers \font\scap=cmcsc10 \hfuzz=0.8truecm
\font\tenmsb=msbm10
\font\sevenmsb=msbm7
\font\fivemsb=msbm5
\newfam\msbfam
\textfont\msbfam=\tenmsb
\scriptfont\msbfam=\sevenmsb
\scriptscriptfont\msbfam=\fivemsb
\def\Bbb#1{{\fam\msbfam\relax#1}}

\def\yup{\overline y}
\def\xup{\overline x}

\null \bigskip  \centerline{\bf THE GAMBIER MAPPING, REVISITED}

\vskip 2truecm
\bigskip 
\centerline{\scap B. Grammaticos}
\centerline{\sl GMPIB (ex LPN), Universit\'e Paris VII} 
\centerline{\sl Tour 24-14, 5$^e$\'etage} 
\centerline{\sl 75251 Paris, France}
\bigskip 
\centerline{\scap A. Ramani} 
\centerline{\sl CPT, Ecole Polytechnique}
\centerline{\sl CNRS, UPR 14} 
\centerline{\sl 91128 Palaiseau, France}
\bigskip 
\centerline{\scap S. Lafortune$^{\dag}$} 
\centerline{\sl LPTM et GMPIB,  Universit\'e Paris VII} 
\centerline{\sl Tour 24-14, 5$^e$\'etage} 
\centerline{\sl 75251 Paris, France}
\footline{\sl $^{\dag}$ Permanent address: CRM, Universit\'e de
Montr\'eal, Montr\'eal, H3C 3J7 Canada}
\bigskip\bigskip

Abstract 
\smallskip \noindent 
We examine critically the Gambier equation and show that it is the generic linearisable equation
containing, as reductions, all the second-order equations which are integrable through linearisation.
We then introduce the general discrete form of this equation, the Gambier mapping, and present
conditions for its integrability. Finally, we obtain the reductions of the Gambier mapping, identify
their integrable forms and compute their continuous limits.
\vfill\eject

\footline={\hfill\folio} \pageno=2

\bigskip
\noindent {\scap 1. Introduction}
\medskip
The classification of the integrable second order differential equations, based on their singularity
properties, resulted to four classes [1]:
\item{-} equations that are simple derivatives of integrable first order equations
\item{-} equations that are autonomous (i.e. they do not have any explicit dependence on the independent
variable) and which are integrable in terms of elliptic functions
\item{-} equations which are non autonomous but in which the the independent variable appears in some simple
form (linearly or at most quadratically) and which define the ${\Bbb P}$ transcendents

\noindent and finally
\item{-} equations which are non autonomous and in which the independent variable enters through some free
functions. These equations are solved by linearisation i.e. they can be converted to a linear differential
system. 

Prominent among this last class is the Gambier equation [2]. This equation is, in fact, the generic
equation of the linearisable class, in the sense that all the others can be obtained as its special
limits. The essential feature of the Gambier equation is that it describes the coupling of two
Riccati equations in cascade (i.e. the solution of the first Riccati equation appears in the
coefficients of the second one). Thus the Gambier equation is best written as: 
$$
\eqalignno{
&y'=-y^2+by+c, & (1.1a) \cr
&x'=ax^2+nyx+\sigma, & (1.1b)
}$$
where $a$, $b$ and $c$ are functions of the independent variable, $\sigma$ is a constant which can
be scaled to $1$ unless it happens to be $0$ and $n$ is an integer. The precise form of the coupling
is indicated by the singularity analysis which, moreover gives constraints on the coefficients $a$,
$b$ and $c$. In fact, out of these three functions  only two (in general) are free. Eliminating $y$
between (1a) and (1b) one can write the Gambier equation as a second order ODE:
$$
\matrix{
\displaystyle{x''={n-1\over n}{x'^2\over x}+a{n+2\over n}xx' +bx' -{n-2\over n}{x'\over
x}\sigma-{a^2 \over n}x^3 + (a'-ab)x^2} \cr
\displaystyle{+ \Big(cn-{2a\sigma\over n}\Big)x-b\sigma-{\sigma^2\over nx}.}
}
\eqno(1.2)
$$

An important remark is in order at this point. The equations of the Painlev\'e/Gam\-bier classification
are usually given in canonical form, which means that all possible transformations of the dependent {\sl
and independent} variables have been used in order to simplify their form. This does not seem to be
done in the case of the Gambier equation. Indeed, as we will show in the next section, a suitable
transformation of the dependent and independent variables allows us to put $b=0$. Thus, the Gambier
equation contains only {\sl two} functions, which moreover are constrained by the integrability
requirement.  

The discretisation of the Gambier system leads naturally to what we have called the Gambier mapping.
In [3] we have proposed such a discretisation which we have studied using the discrete analog of the
singularity analysis, namely the property of singularity confinement. In this paper we propose to
reexamine the discrete form of the Gambier equation and determine its most general expression. Once
this form is established we can proceed to the study of its particular, limiting, forms and propose
expressions for the remaining linearisable discrete equations. For the sake of completeness we calculate,
in the next section, the various limits of the Gambier equation in the continuous case. 

\bigskip
\noindent {\scap 2. The Gambier equation and its various limits}
\medskip

The canonical list of second order equations with the Painlev\'e property is still an unsettled
question. The simplest way out of the dilemma is to adopt the attitude of Gambier [2] who has presented
a minimal list of $24$ equations which contain, in principle all the basic equations. The
remaining ones can be obtained through what in modern parlance would be called Miura transformations.
Among the equations of the Gambier list some belong to the linearisable family. Here they are:
$$
\matrix{
\displaystyle{{\rm (G5)}} & \displaystyle{x''=-3xx'-x^3+q(x'+x^2),}\cr \cr
\displaystyle{{\rm (G13)}}&\displaystyle{x''={x'^2\over x} +q{x'\over x}-q'+rxx'+r'x^2,}\cr \cr
\displaystyle{{\rm (G14)}}&\displaystyle{x''=\Big(1-{1\over n}\,\Big){x'^2\over
x}+qxx'-{nq^2\over(n+2)^2}x^3+{nq' \over n+2}x^2,}\cr \cr
\displaystyle{{\rm (G15)}}&\displaystyle{x''=\Big(1-{1\over n}\,\Big){x'^2\over
x}+f_n(q,r)xx'+\phi_n(q,r)x'-{n-2\over n}{x'\over x}-{nf_n^2\over
(n+2)^2}x^3}\cr \cr
&\displaystyle{+{n(f_n'-f_n\phi_n)\over n+2}x^2+\psi_n(q,r)x-\phi_n-{1\over nx}.} }
\eqno(2.1)
$$
To this list one must, in principle, add the equation
$$
\matrix{
\hskip-7.5cm{\rm (G6)}&\displaystyle{x''=-2xx'+qx'+q'x}
}
$$
which is nothing but the derivative of the Riccati equation. It is easy to show that the Gambier
equation (G15) contains all the previous ones: it is in some sense the general linearisable
equation. Instead of using (G15), which corresponds to $\sigma=1$ in (1.2), we will work with (1.2)
 itself where one can directly see the relation to the coupled Riccati's. 

First we start with (1.2) for $\sigma=1$, and reduce it to its canonical form. For this we introduce the
following transformation of the independent variable $t$ to a new variable $T$ through ${dT\over dt}=g$
where $g$ is defined by ${1\over g}{dg\over dt}=b{n\over n-2}$ and simultaneously $X=gx$. This leads to
an equation of the form (1.2) with $b=0$, which must be considered its canonical form (similarly (G15) is
canonical for $\phi=0$). Moreover the Painlev\'e property requirement introduces one further relation
between $a$ and $c$ (or, equivalently, between $f$ and $\psi$).

Equation (G14) is the easiest to obtain: it suffices to take $\sigma=0$. The canonical form
corresponds to $b=c=0$. Indeed, in addition to the independent variable transformation which allows to
put $b=0$, when $\sigma=0$ we have an additional gauge freedom which allows us to put $c=0$.
$$
x''={n-1\over n}{x'^2\over x}+a{n+2\over n}xx' -{a^2 \over n}x^3 + a'x^2.
\eqno(2.3)
$$
(with $b=c=0$ equation (1.1a) leads to $y={1\over z-z_0}$ and (1.1b) for $\sigma=0$ is reduced to a
linear ODE for $1/x$). 

Equation (G13) requires that we take the limit $n\rightarrow \infty$ on (1.2).
The result is (where $d={\displaystyle \lim_{n\rightarrow \infty}}cn$):
$$
x''={x'^2\over x}+axx' +bx' -{x'\over
x}\sigma + (a'-ab) x^2 + dx-b\sigma.
\eqno(2.4)
$$
Equation (2.4) is (G13) in non canonical form. In order to reduce it to the standard expression we
take $b=0$  and introduce a gauge $x\rightarrow \rho x$ such that
$d=\rho''/\rho-\rho'^2/\rho^2$. The equation reduces then to
$$
x''={x'^2 \over x} + q{x'\over x}-q'+rx'x+r'x^2.\eqno(2.5)
$$  
which is just (G13). What does the limit $n\rightarrow \infty$ really means in the level of the
coupled Riccati's? Since $n$ goes to infinity $y$ must go to zero for the equation to
remain meaningful  and thus the quadratic term in (1.1a) disappears. The canonical form corresponds to
$b=0$ and a new function is introduced through $d\equiv nc$. Finally, if we divide (1.1b) by $x$ and take
the derivative a term $ny'$ appears, which from (1.1a) is equal to $d$. Thus, equation (G13) is, in fact,
nothing but a derivative of a Riccati after we have divided by the dependent variable.  

 Finally in order to obtain (G5) we start by taking $n=1$ which makes the
$x'^2/x$ term vanish. Integrability implies $\sigma=0$ and we choose $a=-1$, $c=0$. This leads to
the equation:
$$
x''=-3xx'-x^3+b(x'+x^2) \eqno(2.6)
$$
which is (G5) in canonical form. Finally it seems that (G6) is not in any sense related to the Gambier
equation (G15).

\bigskip
\noindent {\scap 3. The discrete analog of the Gambier equation, revisited}
\medskip

The discretisation of the Gambier equation is based on the idea of two Riccati equations
in cascade. The discrete form of the first is simply:
$$\yup={ay+b\over y+1}\eqno(3.1)$$
where $y\equiv y_n$ and $\yup\equiv y_{n+1}$. The denominator of (3.1) can be generically be brought
to this form through a scaling of
$y$ and a division by an over-all factor. The second equation which contains the coupling can be
discretised in several, not necessarily equivalent, ways. In [3] we have proposed the
discretisation:
$$\xup={fxy+\sigma\over 1-gx}\eqno(3.2)$$ 
A different approach could be based on the direct discretisation of (2.3) in the form:
$$\xup-x=-fx\xup+(gx+h\xup)y+k\eqno(3.3)$$
In what follows we shall not choose {\sl a priori} a particular form. We shall rather start (in the
spirit of [4]) from a generic coupling of the form:
$$\alpha x\xup y+\beta x\xup+\gamma\xup y+\delta\xup+\epsilon xy+\zeta x+\eta
y+\theta=0\eqno(3.4)$$
Implementing a homographic transformation on $x$ and $y$ we can generically bring (3.4) under
the form:
$$x\xup+\gamma\xup y-\epsilon xy-\theta=0\eqno(3.5)$$
(the sign changes were introduced for future convenience). A choice of different
transformations can bring (3.4) to the form (3.3) while (3.2) can be obtained through a
special choice of the parameters of (3.4). Note that (3.4) contains an `additive' type
coupling $x\xup+\delta\xup+\zeta x+\eta y+\theta=0$ for special values of its parameters, but
the generic form (3.5) is that of a `multiplicative' coupling where $\gamma,\epsilon$ do not
vanish. Solving (3.5) for
$\xup$ we obtain the second equation of the discrete Gambier system in the form:
$$\xup={\epsilon xy+\theta\over x+\gamma y}.\eqno(3.6)$$
Clearly, a scaling freedom remains in equation (3.6). We can use it in order to bring it to the
final form:
$$
\xup={xy/d+c^2 \over x+dy} \eqno(3.7)
$$
  Eliminating $y$ and $\yup$ from (3.1), (3.6) and its upshift, we
can obtain a 3-point mapping for $x$ alone but the analysis is clearer if we deal with both $y$ and
$x$.

The main tool for the investigation of the integrability of the Gambier mapping will be the
singularity confinement criterion [5]. A first remark before implementing the singularity confinement
algorithm is that the singularities of a Riccati mapping are automatically confined. Indeed, if we start
from
$\overline x=(\alpha x+\beta)/(\gamma x+\delta)$ and assume that at some step
$x=-\delta/\gamma$, we find that $\overline x$ diverges  but $\overline {\overline x}$
and all subsequent $x$'s are finite. Thus, the intrinsic singularities of (3.6) do not
play any role. However, the singularities due to $y$ (obtained from (3.1)) may cause
problems at the level of (3.6). Whenever $y$ takes a value that corresponds to either of the
two roots $\pm c$ of the equation:
$$y^2-c^2=0 \eqno(3.8)$$
we obtain $\xup=\pm c/d$  irrespective of the value of $x$ and thus
the variable $x$ loses a degree of freedom. On the other hand, once we enter a singularity
there is no way to exit it unless $y$ assumes again a special value after a certain number of
steps. Thus, if we enter the singularity through, say $y=c$ we can exit it through $y=-c$
after $N$ steps. However, if $y$ were to take the value $c$ again some steps after taking it
for the first time, then it would take it periodically and the singularity would be periodic.
This is  contrary to the requirement that the singularity be movable: a periodic singularity
(with fixed period) is `fixed' in our terminology.

The first singularity condition can thus be obtained in the following way. We assume that at some
step $y$ assumes the value $c$ solution of the condition (3.8). This value of $y_0=c_0$ evolves
under the action of the Riccati and we obtain after $N$ steps, $y_N$. We require that
$$
y_N=-c_N, \eqno(3.9)
$$
i.e. the second root of (3.8). It is thus straightforward to write the first confinement
conditions for the first few values of $N$. We have for instance
$$
\matrix{
\displaystyle{N=1}&\displaystyle{{ac+b \over c+1}+\overline c =0} \cr
\displaystyle{N=2}&\displaystyle{{\overline a (ac+b)+ \overline b (c+1) \over ac+b+c+1}+
\overline{\overline c}=0}
}
\eqno(3.10)
$$
and so on. The equivalent of this requirement in the continuous case is that the resonance be
integer. We see here that the discrete condition is much more complicated and while one can
easily compute the first few instances no general expression can be given.
Once $y$ passes through the second special value $-c$, there is a possibility for $x$ to
recover its lost degree of freedom through an indeterminate form $0/0$. This is the
confinement condition. In full generality (and somewhat abstract form) it reads:
$$x_N+d_N y_N=0\eqno(3.11)$$
or, using (3.9),
$$
x_N=d_Nc_N \eqno(3.12)
$$
where $x_N$ is the $N$-th iterate of $x$ through (3.6). We have for example
$$
\matrix{
\displaystyle{N=1}&\displaystyle{{c \over d}=\overline c \overline d} \cr
\displaystyle{N=2}&\displaystyle{{1\over \overline d}{(ac+b)c+d\,\overline d \,{\overline c}^2(c+1)
\over c(c+1) + d \, \overline d (ac+b)}=\overline{\overline d} \,\overline{\overline c}}.
}\eqno(3.13)
$$
The two confinement conditions put constraints on the coefficients $a$, $b$, $c$ and $d$ just as the
Painlev\'e requirement restricts the parameters in the continuous case. The better approach is to
start with given $a$, $c$ and use (3.10) to solve for $b$. The second condition becomes then an
equation for $d$. For $N=1$ we find explicitely $d \overline d = c/\overline c$. For $N=2$ the
equation for $b$ is linear and the one for $d \overline d$ is just a homographic mapping with
coefficients depending on $a$, $b$, $c$. 

One important question that remains to be addressed is
that of the continuous limit of the Gambier mapping. We start from the system 
$$
\eqalignno{
 &\yup = {ay+b \over y+1}  &(3.18a) \cr
&\xup={xy/d+c^2 \over x+dy} & (3.18b)
}
$$
and introduce the following expansions for the parameters
$$
\matrix{
\displaystyle{a=1+(\alpha+{\gamma'\over \gamma }-{2h\over n})\epsilon,} \cr
\displaystyle{b=\gamma \epsilon^2,} \cr
\displaystyle{c={n\gamma \over 2} \epsilon^2,}\cr
\displaystyle{d=1+\delta\epsilon,} 
} \eqno(3.19)
$$
and for the dependent variables 
$$y={n\gamma \over nY+h}\epsilon \eqno(3.20a)$$
$$x={n\gamma (f X-1)\over 2(f X+1)} \epsilon^2\eqno(3.20b)$$
where $f=\delta +\gamma'/(2\gamma)$ and $h=f'/f$.
We obtain at the limit $\epsilon \rightarrow 0$ the two Riccati's:
$$
\eqalignno{
&X'=-f^2X^2+nXY+1&(3.21a)\cr
&Y'=-Y^2-\alpha Y+\gamma - {\alpha h\over n} +{h^2 \over n^2}-{h'\over n}
& (3.21b)}
$$
where the coefficient of the coupling term $n=2c/b$ is {\sl a priori} a function but with hindsight we
have ignored its derivatives. While the continuous limit takes quite expectedly the form of two
Riccati's in cascade we have still to show that they are indeed of the Gambier form and in particular
that the coefficient of the coupling term $n$ is in fact an integer and equal to $N$. 

The key to this proof is the first confinement condition. Let us start with $y_0=c$. Given the
dependence of $c$ on $\epsilon$ (3.19) we have 
$$y_0={n\gamma \over 2} \epsilon^2\eqno(3.22)$$
In order to do away with the $\epsilon^2$ factor we introduce the auxiliary quantity $\psi$ through
$y=\epsilon^2\psi$ and rewrite (3.22) as:
$$\psi_0={n\gamma \over 2}\eqno(3.23)$$
The confinement condition is 
$$\psi_N=-{n\gamma \over 2}\eqno(3.24)$$
let us now compute $\psi_N$ using the discrete Riccati (3.18) at lowest order in $\epsilon$.
Substituting the expressions (3.19) of $a,b$ we have at lowest order:
$$\overline\psi=\psi+\gamma \eqno(3.25)$$
Thus $\psi_N=\psi+N\gamma $ and substituting the values of $\psi_0$ and $\psi_N$ we find $n=N$. Thus the
coupling coefficient does indeed go over to the integer $N$ which is the number of steps required for
confinement. (Had we started from $y_0=-c$ we would have obtained $n=-N$. The fact that $\pm c$ play
different roles is due to the fact that the discrete Riccati (3.18) is not symmetric with respect to the
upward-downward evolution).

We must remark here that the above continuous limit is incompatible with $N=1$. Indeed for $N=1$
condition (3.13) implies $2\gamma\delta+\gamma'=0$ which would make (3.20) meaningless. This is related
to the fact that in the continuous case $n=1$ is never integrable for $\sigma=1$. In fact for $N=1$ the
only meaningful limit of (3.18) is the one that takes $\sigma$ to zero. It turns out that this limit is
different from (3.20). In fact taking 
$$y={\gamma \over Y}\epsilon \eqno(3.26a)$$
$$x={\gamma ( X-1)\over 2( X+1)} \epsilon^2\eqno(3.26b)$$
and $a=1+(\alpha+{\gamma'\over \gamma })\epsilon$, with $b,c,d$ as in (3.19) with $n=1$, we obtain at the
limit
$\epsilon\to 0$ the system:
$$
\eqalignno{
&X'=XY&(3.27a)\cr
&Y'=-Y^2-\alpha Y+\gamma & (3.27b)}
$$
We remark that in this case the equation for $X$ becomes almost trivial since the $X^2$ term vanishes
together with $\sigma$.

\bigskip
\noindent {\scap 4. Nongeneric forms of the Gambier mapping}
\medskip
An exhaustive study of all the nongeneric cases of the Gambier mapping is a task that lies beyond the
scope of this work. In principle one has to go back to the system (3.1)-(3.4) and, following the steps
of the derivation of (3.5), identify all instances where some transformation cannot be applied. The bulk
of the resulting calculations makes this problem hardly tractable and we prefer, in what follows, to
limit somewhat our scope.

We start thus with the Gambier mapping in its reduced form (3.1)-(3.7) and consider the cases where the
coefficients that have been assumed to be nonvanishing, do vanish. We are thus led to the
systems given below. The equation for $x$ assumes one of the following forms:
$$
\eqalignno{
&\xup={xy/d+c^2 \over x+dy}, & (4.1) \cr
&\xup={xy+c^2 \over x}, & (4.2) \cr
&\xup={xy+c^2 \over y}. & (4.3) 
}
$$while that for $y$ is given by:
$$
\eqalignno{
&\yup={ay+b \over y+1}&(4.4)\cr
&\yup={ay+1 \over y}&(4.5)\cr
&\yup=y+b&(4.6)
}
$$
all the other cases  obtained from (3.1)-(3.7) can be brought to one of the above using homographic
transformations on $x$ and $y$.
Next we shall investigate the singularity confinement property of the system consisting of one of the
(4.1), (4.2), (4.3) coupled to one of the (4.4), (4.5), (4.6). There exist in principle 9 possible
couplings, the one of (4.1) and (4.4) being the full discrete Gambier system studied in the previous
section. 

In order to investigate the coupling (4.1)-(4.5) we apply the singularity confinement method. The
principle is the same as for the full Gambier case: we enter a singularity when $y$ passes through the
value $c$. In order to confine this singularity we require that, after $N$ steps, $y$ pass through $-c$
and moreover $x$ assume an indeterminate form $0/0$. The condition for $y$ to be equal to $-c$ can be
worked out for the first few values of $N$:
$$
\matrix{
\displaystyle{N=1}&\displaystyle{{ac+1 \over c} = -\overline{c},} \cr
\displaystyle{N=2}&\displaystyle{{\overline{a}(ac+1)+c \over ac+1} = 
-\overline{\overline c}.} 
}
\eqno(4.7)
$$
The corresponding conditions for the denominator of $x$ to vanish (which, in view of (4.7), entails the
vanishing of the numerator) read:
$$
\matrix{
\displaystyle{N=1}&\displaystyle{{c \over d} = \overline c \overline d} \cr
\displaystyle{N=2}&\displaystyle{{1 \over \overline d}{c(ac+1)+d\overline d 
{\overline c}^2 c \over c^2 + d\overline d(ac+1)}=\overline{\overline c}\, 
\overline{\overline d}.}
}
\eqno(4.8)
$$
In order to obtain the continuous limit of this system we start with the equation for $y$. The only
continuous limit of (4.5) is obtained for $y=i+\epsilon Y$. However this is {\sl incompatible} with the
integrability condition where $y$ assumes the values $c$ and $-c$ (after $N$ steps). Thus the system
(4.1-5) although integrable as a discrete system does not possess an {\sl integrable} continuous limit.

Next we consider the coupling (4.1)-(4.6) which can be treated just as the previous case. The first few
conditions for the singularity to be confined are:
$$
\matrix{
\displaystyle{N=1}&\displaystyle{c+b = -\overline{c},} \cr
\displaystyle{N=2}&\displaystyle{c+b+\overline b = -\overline{\overline c}.} 
}
\eqno(4.9)
$$
combined with
$$
\matrix{
\displaystyle{N=1}&\displaystyle{{c \over d} = \overline c \overline d} \cr
\displaystyle{N=2}&\displaystyle{{1 \over \overline d}{c(c+b)+{\overline c}^2d\overline d 
  \over c  + d{\overline d}(c+b)}=\overline{\overline c}\, 
\overline{\overline d}.}
}
\eqno(4.10)
$$
In this case the continuous limit is obtained through: $x=\epsilon X$, $c=\epsilon \gamma$,
$d=1+\epsilon \delta$ and $y=Y$. From the constraint (4.9) on $b,c$ we find that at lowest order
we have $b=-2c/N$. The continuous limit is then straightforward, but one must also verify the second
integrability condition (4.10). It turns out that the resulting form is noncanonical. In order to bring
it under canonical form a further transformation is needed on $X$: $X=\gamma(1-\delta W)/(1+\delta W)$
and, moreover, we must take $\gamma'=0$. In the case $N=1$, (where we have from (4.10):
$2\delta+\gamma'/\gamma=0$), the canonical form can be recovered through a simpler transformation,
$X=-\gamma+1/W$, leading to the linear equation: $W''+W'\gamma'/\gamma+W(\gamma'/\gamma)'=0$ which for
$\gamma'=0$ reduces to just $W''=0$.

We turn now to the case of the mapping (4.2) coupled to any of the three homographic for $y$ (4.4-6). A
general remark is in order here. The mapping (4.2) has as only singularity $y=\infty$, i.e. $\xup$ is
defined independently of the value of $x$ only when $y=\infty$. Once $y$ in mappings (4.4) or (4.5)
hits this special value, (4.2) loses one degree of freedom and cannot recover it because $y$ cannot
become infinite again (unless the mapping for $y$ is periodic which we have excluded from the outset).
Thus the combination of (4.2) with either of (4.4) or (4.5) is never integrable. On the contrary (4.2)
coupled to (4.6) is always integrable because the latter, being linear, can never lead to $y=\infty$.
In this case we find at the continuous limit equation (G6). Indeed, putting: $x=1+\epsilon X$,
$y=2+\epsilon^2 Y$ and $b=\beta\epsilon^3,c^2=-1+\gamma \epsilon^2$ we obtain
$X'=-X^2+Y+\gamma$, $Y'=\beta$. Eliminating $Y$ leads to $X''=-2XX'+\beta+\gamma'$ which can be brought
to the canonical form (G6) through a simple translation.

Analogous arguments do apply to the case of the mapping (4.3). The singularity of this mapping occurs
only if $y=0$. Again, the argument of $y$ taking twice the value being possible only if (4.4-6) are
periodic, precludes the integrability of (4.3) coupled to any of these three. However there exists a
case where $y$ {\sl cannot} vanish. This is the case of (4.4) for $b=0$. This is the only integrable
case of mapping (4.3) coupled to (4.4). However it is a trivial one. By transforming $y\to1/y$ both
mappings become linear.
\bigskip
\noindent {\scap 5. Conclusion}
\medskip
In this paper we have examined the Gambier equation in both its continuous and discrete forms. For the
continuous Gambier system we have shown that it is the generic second-order differential linearizable
system: the other second order linearizable ODE's can be obtained as special limits of the Gambier
equation. In the discrete case we have obtained the Gambier mapping starting from the most general
discrete Riccati in cascade system (instead of introducing an {\sl ad hoc} parametrisation as we did in
[3]). This most general form has made possible the interpretation of the number of steps necessary for
confinement. In the particular case of the Gambier mapping this integer coincides with the one appearing
in the coupling term of the ODE's obtained in the continuous limit and which is equal to the resonance
of the Painlev\'e expansion. 

This remark raises two important issues. The first is whether there exists
a systematic relation between the Painlev\'e resonance and the number of steps for confinement, i.e. the
length of the singularity pattern. We believe that the answer is, in general, negative, despite some
tempting results like the Gambier system. The second remark is even more crucial. Since the Gambier
system confines for any number of steps $N$, the limit $N\to\infty$ does in principle exist. We can thus
wonder what is the meaning of confinement that requires an infinite number of steps. How can one
distinguish the $N=\infty$ confining case from a nonconfining one? Although we cannot offer a rigorous
statement, we can present some elements of an answer based on our experience with integrable discrete
systems. In a nonconfining system the analysis of the singularity shows that there is {\sl no}
possibility for confinement ever. In many cases one can even formulate this in rigorous terms and prove
the impossibility of confinement. In the case of a confining mapping the analysis indicates that the
possibility of confinement does exist but is simply delayed (and pushed to infinity at the limit). More
complicated situations may exist, those, among others, involving the discrete derivatives of homographic
mappings. Clearly, at this level the refinement, the notion of confinement itself becomes quite delicate.

The reduced cases of the discrete Gambier system have been only cursorily studied in this work. The
particular case where the Gambier mapping reduces, for $N=1$, to the 2-dimensional projective system was
not contained in the forms studied here. In order to obtain it one must go back to the initial
complete form of the discrete Gambier system and perform the appropriate reductions there. This question
is under active investigation [6].
 \bigskip
\noindent {\scap Acknowledgements}.
\smallskip
\noindent 
 S. Lafortune acknowledges two
scholarships: one from NSERC (National Science and Engineering Research Council of Canada) for his Ph.D.
and one from ``Programme de Soutien de Cotutelle de Th\`ese de doctorat du Gouvernement du Qu\'ebec'' for
his stay in Paris.
\bigskip
{\scap References}
\smallskip 
\item{[1]} E.L. Ince, {\sl Ordinary differential equations}, Dover, New York, 1956.
\item{[2]} B. Gambier, Acta Math. 33 (1910) 1.
\item{[3]} B. Grammaticos and A. Ramani, Physica A 223 (1995) 125.
\item{[4]} A. Ramani, B. Grammaticos and G. Karra, Physica A 181 (1992) 115.
\item{[5]} B. Grammaticos, A. Ramani and V. Papageorgiou, Phys. Rev. Lett. 67 (1991)
1825.
\item{[6]} B. Grammaticos, A. Ramani, K.M. Tamizhamni and S. Lafortune, {\sl Again, linearisable
mappings}, preprint.
 \end